\documentclass[11pt]{article}
\usepackage{latexsym}
\usepackage{epsfig}
\usepackage{a4}
\usepackage{amsfonts}

\makeatletter
\expandafter\let\expandafter\reset@font\csname reset@font\endcsname

\makeatother

\newcommand{\p}{^{\prime}}

\def\be{\begin{equation}}
\def\ee{\end{equation}}
\def\bea{\begin{eqnarray}}
\def\eea{\end{eqnarray}}
\def\dd{\partial}
\def\D{\partial}
\def\'{\prime}
\def\a{\alpha}
\def\b{\beta}

\def\d{\delta}

\def\eps{\varepsilon}
\def\ba{\begin{array}}
\def\ea{\end{array}}

\makeatletter
\def\binrel@#1{\begingroup
  \setboxz@h{\thinmuskip0mu
    \medmuskip\m@ne mu\thickmuskip\@ne mu
    \setbox\tw@\hbox{$#1\m@th$}\kern-\wd\tw@
    ${}#1{}\m@th$}%
  \edef\@tempa{\endgroup\let\noexpand\binrel@@
    \ifdim\wdz@<\z@ \mathbin
    \else\ifdim\wdz@>\z@ \mathrel
    \else \relax\fi\fi}%
  \@tempa
}
\let\binrel@@\relax
\def\overset#1#2{\binrel@{#2}%
  \binrel@@{\mathop{\kern\z@#2}\limits^{#1}}}
\def\underset#1#2{\binrel@{#2}%
  \binrel@@{\mathop{\kern\z@#2}\limits_{#1}}}

\begin{document}

\begin{flushright}
LU-ITP 2005/15
\end{flushright}

\begin{center}
{\LARGE \bf  Symmetric reggeon interaction \\  in perturbative QCD
}

\vspace{1cm}
R. Kirschner
 \vspace{.5cm}

 Institut f\"ur Theoretische Physik, Universit\"at Leipzig,
\\
Augustusplatz 10, D-04109 Leipzig, Germany

\end{center}

\vspace{1cm}
\begin{center}
{\bf Abstract}
\end{center}
\noindent
Integral kernels describing the pair interaction of reggeized gluons and
quarks are reconstructed in terms of conformal symmetric 4-point functions
in the transverse plane of impact parameters.
\vspace{1cm}

\section{Introduction}
\setcounter{equation}{0}

The conformal symmetry emerging in the leading order BFKL equation
\cite{BFKL} played a major role in its solution \cite{Lipatov:1985uk} 
and in a number of related investigations.
The symmetry is useful in formulating the multiple reggeized gluon exchange
exhibiting integrability properties \cite{LevPadua, FK}.
Remarkable symmetry features of the gluonic reggeon interaction have been
pointed out \cite{Lipatov:1993qn} and 
relevant methods of integrable systems have
been adapted to the problem
\cite{DeVega:2001pu,Derkachov:2001yn,Karakhanian:1998gy,Derkachov:2000ne}. 
This symmetry has been observed also in the interaction of reggeized quarks
with each other and with the leading reggeized gluons \cite{Kirschner:1994rq}.

In the BFKL pomeron transition vertices \cite{Bartels:1993ih,Bartels:1994jj} 
the conformal symmetry structure is an important aspect
\cite{Bartels:1995kf,Lotter:1996vk,Vacca:1998kc,Korchemsky:1997fy}.
The role of this symmetry in the dipole picture 
\cite{Nikolaev:1994uu,Mueller:1993rr} and its
relation to the BFKL formulation has been shown e.g. in the papers
\cite{Peschanski:1997yx,Navelet:1997xn,Bialas:1997xp}.

The leading order reggeized gluon interaction and the triple vertex of BFKL
pomerons have been analyzed recently with respect to their conformal
symmetry properties \cite{Bartels:2004ef}, 
discussing also  the relation to the 
dipole picture and the Balitsky-Kovchegov equation. 

In the generalized leading logarithmic approximation the exchanged reggeised
QCD quanta, called reggeons in the following, appear as particles moving in
transverse (impact parameter) space  interacting pairwise.
The symmetry acts as the M\"obius transformation on the (complex valued)
position variable running over the transverse plane,

\be
\label{M"obius}
z \rightarrow z\p = { az+b \over cz+d}, \ \
z \rightarrow z^{* \prime } = { \tilde az + \tilde b \over \tilde cz+
\tilde d} \ \ . 
\ee

In view of the symmetry of the interaction the reggeon states can be
described by holomorphic and anti-holomorphic weights $(\ell ) = (\ell,
\tilde \ell )$ characterizing the conformal representations, or by the
scaling dimension $\ell + \tilde \ell$ and the spin $\ell - \tilde \ell =
[\ell]$. The states of a single leading gluonic reggeon
are characterized by the weights $(\ell) = (0,0)$ and the fermionic reggeons
by $(\ell) = (\frac{1}{2},0)$ or $(0, \frac{1}{2})$ in dependence on their
s-channel helicity. The interaction of fermionic reggeons of opposite
helicity results in double-logarithmic contributions. These contributions
can be accounted for by displacing in this anti-parallel helicity
configuration the weights as $(\ell) = (\frac{\omega}{4}, \frac{1}{2} -
\frac{\omega}{4})$ or $( \frac{1}{2} - \frac{\omega}{4},\frac{\omega}{4} )$
\cite{Kirschner:1994rq}.
Here
$\omega$ is the complex angular momentum i.e. the variable conjugated to
the energy squared $s$ in the Mellin transformation defining the partial
waves in the Regge asymptotics.

The states corresponding to the principal series of representations of
$SL(2, C) $ have integer or half-integer values of spin $n$ and scaling
dimension $1+2 i \nu $ with $\nu$ real,
$(\ell) = ({1+n \over 2} + i \nu, {1-n\over 2} + i \nu ) $.

In the present paper we reconstruct the pairwise reggeon interaction
relying  on this symmetry. We represent the interaction in terms of integral
operators with integrations over the transverse impact parameter plane.
The input are the mentioned conformal weights of the one-reggeon states
and the eigenvalues of the QCD one-loop interaction operators 
on two-reggeon states corresponding to the pricipal series.
The latter are well known for the BFKL  (gluon-gluon) case \cite{BFKL} 
and have been obtained in \cite{Kirschner:1994rq} for the 
cases involving fermionic reggeons by
solving the corresponding version of BFKL-type equations 
in momentum representation in the forward kinematics.

This study was motivated in particular by previous studies concerning the
Bjorken asymptotics, where the one-loop parton interaction or leading-twist
composite operator renormalization has been represented in terms of
conformal symmetric operators \cite{Derkachov:2001km,Kirschner:2004nt}.

\section{Symmetric correlators and operators}
\setcounter{equation}{0}
\subsection{Correlators}

We are going to build the symmetric interaction operators with integral
kernels being symmetric 4-point functions. Symmetric $n$-point functions with
the weights $( \ell_i)$ corresponding to the point $x_i$ obey
\be
\label{sym cond}
\sum_i^n  S_{i,\ell_i}^a \ G(x_1, ...,x_n) = 0, \ \
\sum_i^n  \tilde S_{i,\ell_i}^a \ G(x_1, ...,x_n) = 0.
\ee
The generators of the holomorphic M\"obius transformation (\ref{M"obius})
are
\be
S_{i,0}^- =\D_i,  \ \ S_{i,0}^0 = x_i \D_i, \ \  S_{i,0}^+ = - x_i^2
\D_i, \ee
and the ones of the weight $\ell $ representation are
\be
S_{i,\ell}^- = S_{i,0}^-  , \ \
S_{i,\ell }^0 = x_i^{-\ell} S_{i,0}^0 x_i^{\ell}, \ \
  S_{i,0}^+ =  x_i^{-2\ell } S_{i,0}^+  x_i^{2 \ell }.
\ee
The anti-holomorphic generators $ \tilde S_{i, \ell}^a $ have the
analogous
form with the weight $\ell $ replaced by $\tilde \ell $ and the holomorphic
variables $x_i$ and derivatives $\D_i$ replaced by the anti-holomorphic
ones. The symmetry conditions (\ref{sym cond}) are solved by a 
power-like expression
in the differences of coordinates $x_{ij} = x_i - x_j $
\be
\label{sym corr}
\prod_{i<j} x_{ij}^{a_{ij} } \ \ x_{ij}^{* \tilde a_{ij}}
\ee
with the exponents $(a_{ij} = a_{ji}) $ obeying
\be
\label{sym exp}
\sum_{j=1}^n a_{ij} = - 2 \ell_i, \ \ \
\sum_{j=1}^n \tilde a_{ij} = - 2 \tilde \ell_i .
\ee
For $n=2, 3, 4$ this includes the well-known facts that the 3-point
functions are in general uniquely defined by the weights, that 2-point
functions are non-vanishing for coinciding weights only, and that the
4-point functions are determined by the weights up to an arbitrary function
of the anharmonic ratio. The latter case $n= 4$ is of particular interest in
our context. We denote the index values $i, j$ by $1, 2, 1\p, 2\p $ and the
corresponding weights by $\ell_1, \ell_2, \bar \ell_{1\p} \bar \ell_{2\p} $.
We parametrize the exponents as
\bea
\label{4point exp}
a_{12} = d + {\sigma \over 2} - \ell_1 - \ell_2, \ \
a_{1\p 2\p} = d + {\sigma \over 2} - \bar \ell_{1\p}  - \bar \ell_{2\p}, \cr
a_{12\p} = h + {\sigma \over 2} - \ell_1 - \bar \ell_{2\p}, \ \
a_{1\p 2} = h + {\sigma \over 2} - \bar \ell_{1\p}  - \ell_{2}, \cr
a_{11\p} = -d-h  - \ell_1 - \bar \ell_{1\p},    \ \
a_{2 2\p} = -d -h  - \ell_{2}  - \bar \ell_{2\p}, \cr
\sigma = \ell_1 + \ell_2 + \bar \ell_{1\p} + \bar \ell_{2\p}.
\eea
The analogous relations hold for the exponents $\tilde a_{ij} $ of the
anti-holomorphic powers. Here and in the following expressions typically
consist of holomorphic and anti-holomorphic parts and we follow
\cite{Derkachov:2001yn}
in using the short-hand notations
\bea
\label{notations}
(\alpha) = (\alpha, \tilde \alpha), \ \ \
[x]^{(\alpha)} = x^{\alpha} \cdot x^{* \tilde \alpha }, \cr
[\alpha] = \alpha - \tilde \alpha,       \ \ \
a(\alpha ) = {\Gamma (1- \tilde \alpha ) \over \Gamma (\alpha) }.
\eea
If the entries in the doublet $(\alpha) $ are equal numbers we shall sometimes
write e.g. $(1)$ instead of (1,1).

The expressions (\ref{sym corr}), (\ref{sym exp}, (\ref{4point exp})
define single-valued functions of the complex variables $x_i$ if
$[a_{ij}] $ are integers.

The dependence of the 4-point functions on the two doublets of
parameters $(d), (h)$ enters as exponents of anharmonic ratios,
\be
\label{anharm}
r_{ts}^{-d-h} \cdot r_{tu}^h, \ \ \
r_{ts} = {x_{11\p} x_{22\p} \over x_{12} x_{1\p 2\p} }, \ \ \
r_{tu} = {x_{12\p} x_{21\p} \over x_{12} x_{1\p 2\p} }.
\ee
Because these ratios are dependent,
\be
r_{tu} = r_{ts} -1,
\ee
a generic form of the symmetric 4-point fucntion can be represented by
linear combinations of (\ref{sym corr}), (\ref{4point exp}) with only
one doublet of the parameters $(d), (h) $
varying.

In the particular cases of some exponents $a_{ij} $ being negative integers
besides of the monomials (\ref{sym corr}) we have other expressions obeying
the same symmetry conditions (\ref{sym cond}). For example, if $(a_{11\p}) =
(-1) $ then the expression with the factor $|x_{11\p}^2|^{-1} $ replaced by
$ \d^{(2)} (x_{11\p} ) $ has the same behaviour under conformal
transformations.

\subsection{Operators}

We are going to construct symmetric operators using the symmetric $n$-point
functions as integral kernels.

The point about negative integer exponents mentioned above shows up also
 in the construction of operators. This is illustrated in the simple example
of the 2-point function with $(\ell_1) = (\ell_{1\p}) =  ({1 - \varepsilon
\over 2}) $ acting as the operator kernel on $f(x)$.
\be
\label{1-point operator}
\int {d^2 x_{1\p} \over |x_{11\p}|^{2-2\eps } } f(x_{1\p}) =
{\pi \over \eps } f(x_1) +
\int {d^2 x_{1\p} \over |x_{11\p}|^2 } (f(x_{1\p}) - f ( x_1) ) + {\cal O }
(\eps)
\ee
We find the decomposition rule in the limit,
\be
\label{dec rule}
{1 \over |x_{11\p}|^{2-2\eps } } = { \pi \over \eps} \d ^{(2)} (x_{11\p}) +
{1 \over |x_{11\p}^2|_+} + {\cal O}(\eps ).
\ee
In the limit $\eps \rightarrow 0 $ we have two operators (acting on
one-reggeon wave functions) with the same symmetry properties, one of them
is just the identity operator.

Consider also the operator defined by the 2-point function with
$(\ell_1) = (\ell_{1\p}) = (-1 + a +\frac{n}{2}, -1 +a - \frac{n}{2} )$,
$n$ integer.
Acting on the function of the form $ \psi_{b,m} (x) =
x^m   | x |^{-2 + 2b -m} $, $m$ integer,
one obtains a similar function with shifted parameters, $\psi_{a+b, m+n} $,
\bea
\int d^2 x_{1\p}  { x_{11\p}^n \over |x_{11\p}|^{2-2a + n } }
  { \ x_{1\p}^m \over |x_{1\p}|^{2-2b + m }} \ 
= \ \pi \   { \ x_{1}^{n+ m} \over |x_{1}|^{2-2(a+b) +n+ m } } \cr
{ \Gamma (1-a-b + \frac{n+m}{2} ) \ \Gamma (a + \frac{n}{2} ) \ 
\Gamma (b + \frac{m}{2}) \over
\Gamma (1- a + \frac{n}{2} )  \ \Gamma (1- b + \frac{m}{2}) \ 
\Gamma (a + b + \frac{n+m}{2} ) } .
\eea
The operator is defined by the integral for $a > 0, b> 0, a+b < 1$. For
other values of the parameters we define the operator by analytic
continuation provided by the right-hand side.
In terms of the short-hand notation (\ref{notations}) the latter
relation can be written as
\be
\int d^2 x_{1\p} \ [x_{11\p}]^{(-\alpha )} \ [x_{1\p} ]^{-(\beta )} =
\pi  a(\alpha) \ a(\beta) \ a(2 -\tilde \alpha - \tilde \beta ) \
[x_1]^{(-\alpha-\beta +1 )}
\ee
with $(\alpha) = ( 1 - a + \frac{n}{2}, 1 - a - \frac{n}{2}),
(\beta ) = ( 1 -b + \frac{m}{2}, 1 -b -\frac{m}{2}) $.
A more general form of this relation is obtained by shifting the points
$x_1, x_{1\p} $ by $x_2$ and doing the inversion about an arbitrary point
$x_3$.
\bea
\label{masterint}
\int d^2x_{1\p} \ [x_{11\p} ]^{(- \a)} \ [x_{21\p} ]^{(-\b)} \ [x_{31\p}
]^{(- \gamma) }
 \ = \cr  \pi a(\alpha) \ a(\beta ) \ a(\gamma)  \ [x_{12} ]^{(\gamma
-1)} \ [x_{23} ]^{(\alpha -1)} \ [x_{31} ]^{(\beta -1)}, \cr
(\alpha + \beta + \gamma) = (2)
\eea

We turn to the operators acting on two-reggeon states, i.e. on functions of
two complex variables $x_{1}, x_{2} $, the behaviour under conformal
transformations of which is characterized by the weights
$(\ell_1),(\ell_2)$.  Two-reggeon states of definite weight $(\ell_0)$ are
represented by the 3-point functions
\be
\label{3-point}
E_{(\ell_1),(\ell_2)}^{(\ell_0)} (x_1,x_2;x_0) =
[x_{12}]^{(\ell_0-\ell_1 -\ell_2)} \
[x_{10}]^{(\ell_2-\ell_1 -\ell_0)} \
[x_{20}]^{(\ell_1-\ell_2 -\ell_0)}.
\ee
The position variable $x_0$ serves as a label of states within the
representation $(\ell_0)$.

We consider operators acting  symmetrically from
the tensor product representation $V_{(\ell_{1\p})} \otimes
V_{(\ell_{2\p})} $
to $ V_{(\ell_1)} \otimes V_{(\ell_2)} $ . Then their kernels are 
symmetric 4-point
functions (\ref{sym corr}), (\ref{sym exp}) with the weights $(\ell_1),
(\ell_2),
(\bar \ell_{1\p}) = (1- \ell_{1\p}), (\bar \ell_{2\p}) = (1- \ell_{2\p}) $.

$E_{(\ell_1),(\ell_2)}^{(\ell_0)} (x_1,x_2;x_0) $ are eigenfunctions of
these operators in the case $(\ell_1) = (\ell_{1\p}), (\ell_2 ) =
(\ell_{2\p}) $.  Generic symmetric operators with these weights can be
represented in terms of these kernels by variation of only one doublet of
parameter out of the two $(d), (h)$.
If these parameters are related by $ (d+h+1) = (0) $ at
$(\ell_1) = (\ell_{1\p}), (\ell_2 ) = (\ell_{2\p}) $
the integrand simplifies by vanishing of the exponents of $x_{11\p},
x_{22\p} $. In this particular case we denote the kernel by
$K_{(\ell_1),(\ell_2)}^{(d)} (x_1,x_2;x_{1\p}, x_{2\p} )$ and the
corresponding operator by $\hat K_{(\ell_1),(\ell_2)}^{(d)} $.
The eigenvalue relation reads
\bea
\label{eigenrelation}
\hat K_{(\ell_1),(\ell_2)}^{(d)}
E_{(\ell_1),(\ell_2)}^{(\ell_0)} (x_1,x_2;x_0) = \cr
{1 \over \pi^2 } \int d^2 x_{1\p}  d^2 x_{2\p}
K_{(\ell_1),(\ell_2)}^{(d)} (x_1,x_2;x_{1\p}, x_{2\p} )
E_{(\ell_{1\p}),(\ell_{2\p})}^{(\ell_0)} (x_1,x_2;x_0) = \cr
\lambda_{(\ell_1),(\ell_2),(\ell_0)}^{(d)} \
E_{(\ell_1),(\ell_2)}^{(\ell_0)} (x_1,x_2;x_0)
\eea
The eigenvalues are calculated by applying (\ref{masterint})
(compare  \cite{Derkachov:2001yn})
\be
\label{eigenvalues}
\lambda_{(\ell_1),(\ell_2),(\ell_0)}^{(d)} \
= (-1)^{[d+\ell_1 - \ell_2]} \
a(1+d-\ell_1 + \ell_2) a( 1+d+ \ell_1 - \ell_2) \ \
a(1- d - \ell_0 ) a (-d + \ell_0 ).
\ee
Notice that the product $\lambda_{(\ell_1),(\ell_2),(\ell_0)}^{(d)} \
\lambda_{(\ell_1),(\ell_2),(\ell_0)}^{(-d)} $ does not depend on
$ (\ell_0) $. Therefore the product of the corresponding operators is the
identity up to normalization (on the space of functions spanned by
(\ref{3-point}) with $(\ell_0) $ taking the values of the principal series),
\be
\hat K_{(\ell_1),(\ell_2)}^{(d)} \ \hat K_{(\ell_1),(\ell_2)}^{(-d)} \
= \rho(\ell_1, \ell_2, d) \ \hat I.
\ee
Moreover, $ \hat K_{(\ell_1),(\ell_2)}^{(d)} $ obeys the Yang-Baxter
relation with $(d)$ playing the role of the spectral parameter
\cite{Derkachov:2001yn}.

We expand the $(\ell_0)$ dependent factors in the
eigenvalues at $(d) = (\Delta) - \eps$,
\bea
\label{expand}
a(1-\Delta - \ell_0 -\eps) \ a(- \Delta + \ell_0 - \eps) =
a(1-\Delta - \ell_0 ) \ a(- \Delta + \ell_0 ) \cdot \cr
\left \{ 1 + \eps \ [ \ \chi_{-\Delta} (\ell_0 (1- \ell_0) ) +
\chi_{\tilde \Delta} ( \tilde \ell_0 (1- \tilde \ell_0 )) - 4 \psi(1) ]
+ {\cal O} (\eps^2)  \right \}
\eea
Here we have used the notation
\be
\label{chi}
\chi_{\Delta} (x) = \psi(1) - \psi(x + \Delta) - \psi(1-x + \Delta )
\ee
which appeared in \cite{Kirschner:1994rq} in writing the one-loop results for the
eigenvalues of the QCD reggeon interactions with $(\Delta)$ taking the
particular values $(\Delta) = \pm (\ell_1 - \ell_2) $.

We notice further that the factors in the eigenvalues (\ref{eigenvalues})
that do not depend on $(\ell_0)$ have a pole in $\eps$ at
$(d) = \pm (\ell_1 - \ell_2) -  \eps$. This singularity is caused by
some exponents in the kernel approaching negative integers, indeed
\be
(a_{12\p} ) = (-1-d - \ell_1+\ell_2 ), \ \ (a_{21\p} ) = (-1 -d +
\ell_1 - \ell_2).
\ee
From the above discussion it is clear that in this case the definition of
regular symmetric operators takes some care, following the
example of a one-point operator (\ref{1-point operator}) (\ref{dec rule}).
This will be done for the physically relevant cases in the following.

\section{Identical reggeons}
\setcounter{equation}{0}

Consider now the case $(\ell_1-\ell_2) = (0) $ relevant for the
interaction of two gluonic reggeons (BFKL), $(\ell_1) = (\ell_2) = (0)
$,
or of two fermionic reggeons of parallel helicity,
$(\ell_1) = (\ell_2) = (\frac{1}{2}, 0)$ or $(0, \frac{1}{2} )$.
At $(d) = -\eps $ we have
\be
\label{expand0}
\lambda_{(\ell_1),(\ell_2),(\ell_0)}^{(-\eps)} \
= {(-1)^{[\ell_0]} \over \eps^2 }
\left \{ 1 + \eps \ [ \ \chi_{0} (\ell_0 (1- \ell_0) ) +
\chi_{0} ( \tilde \ell_0 (1- \tilde \ell_0 ))  ]
+ {\cal O} (\eps^2)  \right \}
\ee
The coefficient of $\eps^{-1}$ is proportional to the well-known BFKL
pomeron eigenvalues \cite{BFKL,Lipatov:1985uk} 
with $\ell_0 = \frac{1-n}{2} +i \nu, [\ell_0] = n $.
The exponents of the kernels (\ref{4point exp})
at $(d) = (-\eps) $ are
\be
\label{exp0}
(a_{12\p})= (a_{21\p}) = (-1+ \eps ), \
(a_{1\p 2\p}) = (-1-\eps + \ell_1 +\ell_2), \
(a_{12}) = (1 - \eps -\ell_1 -\ell_2 ).
\ee
Consider first the case $
(\ell_1) = (\ell_2) = (\ell) \not= 0 $, where the singularity of the type
(\ref{1-point operator} ) (\ref{dec rule}) appears twice. In the limit $\eps
\rightarrow 0 $ the 4-point function (\ref{sym corr}) and the ones with
some of the negative integer powers replaced by $\d$-functions behave
equally under conformal transformations. We find combinations of these
4-point functions that define regular operators.

Replacing both $|x_{12\p}^2|^{-1} $ and $|x_{21\p}^2|^{-1}$ by the
corresponding $\d$-functions we obtain the permutation operator
$\hat P_{12}$ with the kernel $\d^{(2)} (x_{12\p}) \  \d^{(2)} (x_{21\p}) $.
Replacing instead only one of these negative-integer powers by the
corresponding $\d$-function we define the symmetric operator $\hat
K_{(\ell_1)}^{0 +} $ with the kernel
\bea
\label{kernel0+}
K_{(\ell)}^{0 +} = [x_{12}]^{(1-2 \ell)} \ [x_{1\p 2\p}]^{(-1+ 2 \ell)}
\cr
 \left \{ {1 \over |x_{21\p}^2|_+} \d^{(2)} (x_{12\p}) +
 {1 \over |x_{12\p}^2|_+} \d^{(2)} (x_{21\p}) \right \}
\eea
Approaching $(d) = (0) $ with the generic operator $\hat
K^{(-\eps)}_{(\ell),(\ell)} $ we have the expansion in terms of the 
latter two operators,
\bea
\label{dec0}
\hat K^{(-\eps)}_{(\ell ),(\ell)} = A(\eps)  \hat P_{12} + B(\eps)
K_{(\ell)}^{0 +},  \cr
A(\eps) = \eps^{-2} ( \pi^2 + {\cal O} (\eps) ), \ \ \
B(\eps) = \eps^{-1} ( \pi + {\cal O} (\eps) ).
\eea
We identify the operator $\hat P_{12} \cdot \hat K_{(\ell)}^{0 +} $ as
the one representing the identical reggeon interaction for
$(\ell_1) = (\ell_2) = (\ell) \not= 0 $,
in particular the parallel helicity fermionic reggeons, Fig 1a. 
Its kernel is obtained from
(\ref{kernel0+}) by interchanging the labels $1\p, 2\p $. Its
eigenvalues on the two-reggeon functions (\ref{3-point}) are
\be \label{eigen0}
\chi_{0}
(\ell_0 (1- \ell_0) ) + \chi_{0} ( \tilde \ell_0 (1- \tilde \ell_0 ))
\ee

\begin{figure}[htb]
\begin{center}
\epsfig{file=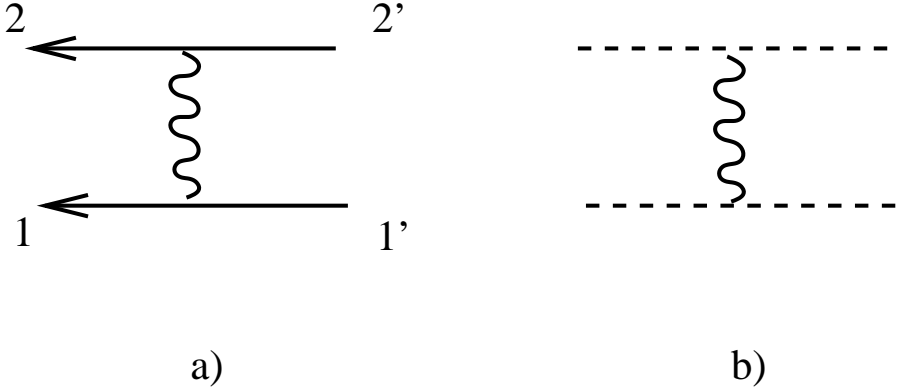,width=12cm}
\end{center}
\vspace*{0.5cm}
\caption{\small
Identical reggeon interaction. a) fermions of parallel helicity, b)
gluonic reggeons (dashed lines). }
\end{figure}

In the case
$(\ell_1) = (\ell_2) = (0) $ simultaneously with $(a_{12\p}) =
(a_{21\p}) $ also $(a_{1\p 2\p})$ approaches $(-1)$. On two-reggeon wave
functions vanishing at coinciding points $x_1 = x_2$ a regular operator
can be defined by a modified combination of the ones above obtained by
the replacement of the powers
$|x_{12\p}^2|^{-1} $ and $|x_{21\p}^2|^{-1}$ by $\d$-functions.
The corresponding kernel can be
written in terms of an integral over the auxiliary point $x_{3\p}$ as
\bea
\label{kernel0+0}
K_{(0)}^{ 0 +} = \int { d^2 x_{3\p} |x_{12}^2| \over |x_{13\p}^2|
|x_{23\p} ^2| } \cdot \cr
\left ( \d^{(2)} (x_{21\p}) \ \d^{(2)} (x_{2\p 3\p})  \ + \
\d^{(2)} (x_{13\p}) \ \d^{(2)} (x_{1\p 3\p} ) \ - \
\d^{(2)} (x_{12\p}) \ \d^{(2)} (x_{2 1\p } ) \right ).
\eea
With the operator defined by this kernel the expansion (\ref{dec0})
holds analogously. The operator
 $\hat P_{12} \cdot \hat K_{(0)}^{0 +} $ represents the gluonic reggeon
interaction, Fig. 1b. Its kernel is obtained from (\ref{kernel0+0}) by
interchanging the labels $1\p, 2\p $ and its eigenvalues are given by
(\ref{eigen0}).

The resulting kernel is the one known in the dipole picture of BFKL
\cite{Mueller:1993rr}. This representation of the BFKL kernel and its relation
to other ones has been studied recently \cite{Bartels:2004ef}.

\section{Fermion - gluon interaction}
\setcounter{equation}{0}

The pair interaction of fermionic and gluonic reggeons in an overall
colour singlet exchange is determined by a symmetric operator with the
weights $(\ell_1) = (\frac{1}{2} , 0), (\ell_2) = (0,0) $. The
corresponding kernel $K_{(\ell_1), (\ell_2)}^{(d)}$ is single-valued for
$[d] = \frac{1}{2} + m $, $m$ integer, and the wave functions
(\ref{3-point}) are single-valued for $[\ell_0] = \frac{1}{2} + n$,
$n$ integer.
At $(d) = (\ell_1 - \ell_2 -\eps) $ the eigenvalues (\ref{eigenvalues})
behave like
\bea
\label{expand1/20+}
\lambda_{(\ell_1),(\ell_2),(\ell_0)}^{(\ell_1 - \ell_2 -\eps)} \
= - {(-1)^{n} \over \eps^2 } (\ell_0 - \frac{1}{2})  \cr
\left \{ 1 + \eps \ [ \ \chi_{-\frac{1}{2}} (\ell_0 (1- \ell_0) ) +
\chi_{0} ( \tilde \ell_0 (1- \tilde \ell_0 ))  +1 ]
+ {\cal O} (\eps^2)  \right \}
\eea
and at $(d) = (-\ell_1 + \ell_2 -\eps) $ we obtain
\bea
\label{expand1/20-}
\lambda_{(\ell_1),(\ell_2),(\ell_0)}^{(-\ell_1 + \ell_2 -\eps)} \
=  {(-1)^{n} \over \eps } (\ell_0 - \frac{1}{2} )^{-1}  \cr
\left \{ 1 + \eps \ [ \ \chi_{\frac{1}{2}} (\ell_0 (1- \ell_0) ) +
\chi_{0} ( \tilde \ell_0 (1- \tilde \ell_0 ))   ]
+ {\cal O} (\eps^2)  \right \}.
\eea

At $(d) = (\ell_1 - \ell_2 -\eps) $ we encounter two exponents (\ref{4point
exp})
approaching negative integers, $ (a_{12\p} ) = (-2, -1) + (\eps), \ 
(a_{12\p}) = (-1, -1) + (\eps) $. 
We have the regular symmetric operators defined by the kernels
\bea
\label{kernel1/20+}
K^{+(\frac{1}{2},0), \d} = - x_{12} \ \ \dd_{2\p} \ \d^{(2)} (x_{12\p})  
\cdot \d^{(2)} (x_{21\p}), \cr
K^{+(\frac{1}{2},0), +} = |x_{12}^2| {x_{1\p 2\p} \over |x_{1\p2\p}^2| }
\left \{
\dd_{2\p} \d^{(2)} (x_{12\p}) {1 \over |x_{21\p}^2|_+}  \ + \
\dd_{2\p} {1 \over |x_{12\p}^2|_+}
 \d^{(2)} (x_{21\p})
\right \},
\eea
and the decomposition
\bea
\label{dec1/20+}
\hat K_{(\frac{1}{2},0),(0,0)}^{(\frac{1}{2}-\eps,-\eps)}
= A_+(\eps) \hat K^{(\frac{1}{2},0),\d} \ + \ B_+(\eps) \
 \hat K^{(\frac{1}{2},0),+}, \cr
 A_+(\eps) = {1\over \eps^2} (\pi^2+ {\cal O} (\eps), \ \
 B_+(\eps) = {1\over \eps} (\pi + {\cal O} (\eps).
\eea
At $(d) = (-\ell_1 + \ell_2 -\eps) $ we have the following exponents
approaching negative integers, $ (a_{12\p} = (a_{1\p 2\p}) = (-1 +
\eps) $.
On wave functions vanishing at coinciding arguments, $x_1 = x_2$, we
have regular symmetric operators given by the kernels
\bea
\label{kernel1/20-}
K^{-(\frac{1}{2},0), \d} \ =  x_{12}^* \ |x_{1\p 2\p}^2|^{-1}
\ \d^{(2)} (x_{12\p})  \ {x_{21\p}\over |x_{21\p}^2| }, \cr
K^{-(\frac{1}{2},0), +} = \ x_{12}^*  \ |x_{1\p2\p}^2|^{-1}
\ {1 \over |x_{12\p}^2|_+ }   \ {x_{21\p} \over |x_{21\p}^2|},
\eea
and the decomposition
\bea
\label{dec1/20-}
\hat K_{(\frac{1}{2},0),(0,0)}^{(-\frac{1}{2}-\eps,-\eps)}
= A_-(\eps) \hat K^{-(\frac{1}{2},0),\d} \ + \ B_-(\eps) \
 \hat K^{-(\frac{1}{2},0),+}, \cr
 A_-(\eps) = {1\over \eps} (\pi + {\cal O} (\eps), \ \
 B_-(\eps) =  (1+ {\cal O} (\eps).
\eea

\begin{figure}[htb]
\begin{center}
\epsfig{file=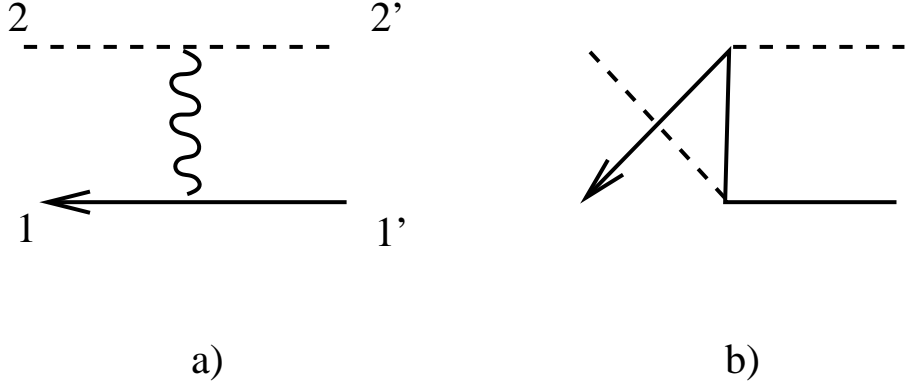,width=12cm}
\end{center}
\vspace*{0.5cm}
\caption{\small
Interaction of reggeized fermion and gluon, a) by gluon exchange, b)
by fermion exchange. }
\end{figure}

 The operator composed out of the above ones as
\be
\label{fg operator}
- \left (
 \hat K^{(-\frac{1}{2},0),\d} \
 \hat K^{(\frac{1}{2},0),+}  \ + \
 \hat K^{(+\frac{1}{2},0),\d} \
 \hat K^{(\frac{1}{2},0),-} \right )
\ee
has the eigenvalues on the two-reggeon wave functions (\ref{3-point})
\be
\label{eigen1/20}
2 \chi_0 (\tilde \ell_0 (1-\tilde \ell_0)) \ + \
 \chi_{\frac{1}{2} }  (\ell_0 (1- \ell_0)) \ + \
 \chi_{-\frac{1}{2}}  (\ell_0 (1- \ell_0)) \ + \ 1
\ee
and in this way represents (up to normalization) the fermion-gluon
reggeon interaction mediated by an $s$-channel gluon, Fig. 2a.

We notice also that the operator
$ \hat K^{(-\frac{1}{2},0),\d} $ having the eigenvalues on the
two-reggeon wave functions (\ref{3-point}) for $\tilde \ell_0 = \ell_0 +
\frac{1}{2} + n $ equal to $(-1)^n (\ell_0 -\frac{1}{2})^{-1} $
represents the fermion-gluon interaction mediated by a fermion, Fig. 2b.

\section{Fermions of anti-parallel helicities }
\setcounter{equation}{0}

Consider first the formal case $(\ell_1) = (\frac{1}{2}, 0), (\ell_2) =
(0, \frac{1}{2}) $, ignoring double-log contributions.
At $(d) = (\pm \ell_1 \mp \ell_2 -\eps) $ the eigenvalues
(\ref{eigenvalues}) are
\bea
\label{expand1/21/2}
\lambda_{(\ell_1),(\ell_2),(\ell_0)}^{(\pm \ell_1 \mp \ell_2 -\eps)} \
=  {1 \over \eps } {\Gamma (\tilde \ell_0\mp \frac{1}{2} ) \Gamma(1-
\tilde \ell_0 \mp\frac{1}{2} ) \over \Gamma(1- \ell_0 \mp \frac{1}{2})
\Gamma(\ell_0 \mp\frac{1}{2}) }    \cr
\left \{ 1 + \eps \ [ \ \chi_{\mp \frac{1}{2}} (\ell_0
(1- \ell_0) ) + \chi_{\mp \frac{1}{2}} ( \tilde \ell_0 (1- \tilde \ell_0
)) +1 ] + {\cal O} (\eps^2)  \right \}
\eea
At $(d) = (\pm \ell_1 \mp \ell_2  ) = (\pm \Delta) $ the exponent
$(a_{21\p})$ or $(a_{12\p})$ turns to $(-1)$ and we have the
regular symmetric operator kernels obtained from
$K_{(\ell_1),(\ell_2)}^{(d)} $ by replacing the factor with the exponent
$-1$, $|x_{ij}^2|^{-1} $, by $\d^{(2)} (x_{ij})$ or by $|x_{ij}^{-2}|_+$.
In this way we are led to define the kernels
\bea
\label{kernel1/21/2}
K^{\frac{1}{2}, \d} = x_{12} \ {x_{1\p 2\p} \over |x_{1\p 2\p}|^2 } \ \ 
\d^{(2)} (x_{21\p}) \ {x_{12\p}^{* 2} \over |x_{12\p}^4| }, \cr
K^{\frac{1}{2}, +} = x_{12} \ {x_{1\p 2\p} \over |x_{1\p 2\p}|^2 } \ \ 
{1 \over |x_{21\p}^2|_+} \ {x_{12\p}^{* 2} \over |x_{12\p}^4| }, \cr
K^{-\frac{1}{2}, \d} = x_{12}^* \ {x_{1\p 2\p}^* \over |x_{1\p 2\p}|^2 } \ \ 
\d^{(2)} (x_{12\p}) {x_{21\p}^2 \over |x_{21\p}^4| }, \cr
K^{-\frac{1}{2}, +} = x_{12}^* \ {x_{1\p 2\p} \over |x_{1\p 2\p}|^2 } \ \ 
{1 \over  |x_{12\p}^2|_+} \ {x_{21\p}^2 \over |x_{21\p}^4| }.
\eea
The operator
$\hat K_{(\ell_1), (\ell_2)}^{(d)} $, with 
$ (\ell_1) = (\frac{1}{2},0), (\ell_2)= (0,\frac{1}{2})$ 
decomposes at $(d) = \pm (\frac{1}{2}, -\frac{1}{2}) - (\eps) $ as
\bea
\label{dec1/21/2}
\hat K_{(\frac{1}{2},0),(0,\frac{1}{2})}^{(\pm \frac{1}{2}-\eps,
\mp \frac{1}{2} -\eps)} = A (\eps) \hat K^{\pm \frac{1}{2},\d} \ + \
B(\eps) \
 \hat K^{\pm \frac{1}{2} +} \ \ ,  \cr
 A (\eps) = {1\over \eps} (\pi + {\cal O} (\eps), \ \
 B(\eps) =  (1+ {\cal O} (\eps).
\eea
The operator composed out of the ones defined by the above kernels 
(\ref{kernel1/21/2}) as
\be
\label{operatorff}
\hat K^{-\frac{1}{2},\d} \
 \hat K^{+\frac{1}{2} +}  \ + \
 \hat K^{+\frac{1}{2}, \d} \
 \hat K^{-\frac{1}{2}, +}
\ee
has the eigenvalues on the two-reggeon wave functions (\ref{3-point})
\be
\label{eigen1/21/2}
 \chi_{-\frac{1}{2}} (\tilde \ell_0 (1-\tilde \ell_0)) \ + \
 \chi_{\frac{1}{2}} (\tilde \ell_0 (1-\tilde \ell_0)) \ + \
 \chi_{\frac{1}{2}}  (\ell_0 (1- \ell_0)) \ + \
 \chi_{-\frac{1}{2}}  (\ell_0 (1- \ell_0)) \ + \ 2.
\ee
It describes in leading $\ln s$ accuracy the interaction of
anti-parallel helicity fermionic
reggeons, Fog. 3,  besides of the two-reggeon states with $[\ell_0] = 0 $, where
double-log contributions appear.

\begin{figure}[htb]
\begin{center}
\epsfig{file=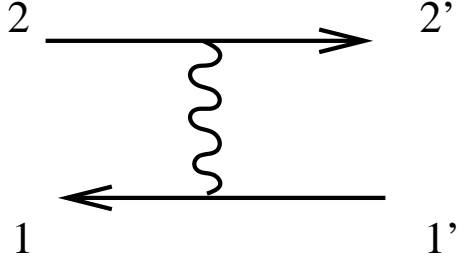,width=6cm}
\end{center}
\vspace*{0.5cm}
\caption{\small
Interaction of reggeized fermions of anti-parallel helicities. }
\end{figure}

In the case $(\ell_1) = (\frac{1}{2}-  \frac{\omega}{4}, \frac{\omega}{4}), 
(\ell_2) = (\frac{\omega}{4}, \frac{1}{2} - \frac{\omega}{4})$,
appropriate for accounting the double-logarithmic contributions in the
anti-parallel helicity fermion exchange, we encounter difficulties
with the formulation of integral operators and wave functions on the plane.
The corresponding kernels (\ref{sym corr}) (\ref{4point exp})
and wave functions (\ref{3-point}) with these weights are not single valued
irrespective to the particular choices of $(d)$ and $(\ell_0)$.

We propose to describe the operators, wave functions and eigenvalues with
these weights determined by the complex angular momentum $\omega $
as the analytic continuation from a series of corresponding objects with 
weights determined instead by even non-positive  integers $-2m$,
$ - 2m \ \ (m= 0, 1 ...) \ \rightarrow \ \omega$,
\bea
\label{ellm}
(\ell_{1, m}) = (\frac{1+m}{2}, - \frac{m}{2} ), \ \ \  
(\ell_{2, m}) = (- \frac{m}{2}, \frac{1+m}{2} ), \cr
(\ell_{1, m} - \ell_{2, m}) =  (\Delta_m ) = 
(\frac{1}{2} +m, -\frac{1}{2} -m ), \ \ 
[\ell_{1, m} - \ell_{2, m}] = 1+ 2 m, \ \ \cr
(\ell_{1, m} + \ell_{2, m} ) = (\frac{1}{2}, \frac{1}{2} ). 
\eea
The eigenvalues (\ref{eigenvalues}) decompose at $(d) = \pm (\Delta_m - \eps) $
as
\bea
\label{expandm}
\lambda_{(\ell_{1,m}),(\ell_{2, m}),(\ell_0)}^{(\pm \Delta -\eps)} \
=  {1 \over \eps } {1 \over 1 + 2m}  
{\Gamma (\tilde \ell_0\mp (\frac{1}{2} +m) ) \Gamma(1-
\tilde \ell_0 \mp(\frac{1}{2} + m) ) \over \Gamma(1- \ell_0 \mp (\frac{1}{2}
+m))
\Gamma(\ell_0 \mp(\frac{1}{2})+m) }    \cr
{\large  \{ } 1 + \eps \ [ \ \chi_{\mp (\frac{1}{2} +m}) (\ell_0
(1- \ell_0) ) + \chi_{\mp (\frac{1}{2} +m } ( \tilde \ell_0 (1- \tilde \ell_0
)) +  \cr  
\psi(1+ 2m) + \psi(2+ 2m) -2 \psi(1) ] + {\cal O} (\eps^2)   {\large \} }
\eea
We see that there is no obstacle to an analytic continuation 
of the expression on the right-hand side.  

We calculate the exponents (\ref{4point exp}) with these weights and for 
$(d) = \pm (\Delta_m - \eps) $ and define kernels in analogy to
(\ref{kernel1/21/2}),
\bea
\label{kernelm}
K^{\frac{1}{2}+m, \d} = x_{12} \left ( {x_{12} \over x_{12}^*} \right )^m
 \ {x_{1\p 2\p} \over |x_{1\p 2\p}^2 | } 
\left ( {x_{1\p2\p} \over x_{1\p2\p}^*} \right )^m\ \ 
\d^{(2)} (x_{21\p}) \ {1 \over |x_{12\p}^2| }
\left ( {x_{12\p} \over x_{12\p}^*} \right )^{-2m-1} , \cr
K^{\frac{1}{2}+m, +} = x_{12} \left ( {x_{12} \over x_{12}1^*} \right )^m
 \ {x_{1\p 2\p} \over |x_{1\p 2\p}^2 |} 
\left ( {x_{1\p2\p} \over x_{1\p2\p}^*} \right )^m\ \ 
{1 \over |x_{21\p}^2|_+}           \ {1 \over |x_{12\p}^2| }
\left ( {x_{12\p} \over x_{12\p}^*} \right )^{-2m-1} , \cr
K^{-\frac{1}{2}+m, \d} = x_{12}^* \left ( {x_{12} \over x_{12}^*} \right
)^{-m}
 \ {x_{1\p 2\p}^* \over |x_{1\p 2\p}^2 | } 
\left ( {x_{1\p2\p} \over x_{1\p2\p}^*} \right )^{-m}\ \ 
\d^{(2)} (x_{21\p}) \ {1 \over |x_{12\p}^2| }
\left ( {x_{12\p} \over x_{12\p}^*} \right )^{2m+1} , \cr
K^{\frac{1}{2}+m, +} = x_{12}^* \left ( {x_{12} \over x_{12}^*} \right )^{-m}
 \ {x_{1\p 2\p}^* \over | x_{1\p 2\p}^2 | } 
\left ( {x_{1\p2\p} \over x_{1\p2\p}^*} \right )^{-m}\ \ 
{1 \over |x_{21\p}^2|_+}           \ {1 \over |x_{12\p}^2| }
\left ( {x_{12\p} \over x_{12\p}^*} \right )^{2m+1}.  
\eea
We have the decomposition of the operators $
\hat K_{(\ell_{1,m}),(\ell_{2, m})}^{(\pm \Delta -\eps)} $ in to the
ones with the above kernels in analogy to (\ref{dec1/21/2}). The series
of operators 
\be
\label{operatorffm}
\hat K^{-(\frac{1}{2}+m),\d} \
 \hat K^{+(\frac{1}{2}+m) +}  \ + \
 \hat K^{+(\frac{1}{2}+m), \d} \
 \hat K^{-(\frac{1}{2}+m), +}
\ee
has the eigenvalues on the two-reggeon wave functions (\ref{3-point})
correspondingly with the weights $\ell_{1, m}) = (\frac{1+m}{2}, - \frac{m}{2} ), \ \ \  
\ell_{2, m}) = (\frac{m}{2}, \frac{1+m}{2} ), (\ell_0) $
\bea
\label{eigenm}
 \chi_{-\frac{1}{2}-m} (\tilde \ell_0 (1-\tilde \ell_0)) \ + \
 \chi_{\frac{1}{2}+m} (\tilde \ell_0 (1-\tilde \ell_0)) \ + \
 \chi_{\frac{1}{2}+m}  (\ell_0 (1- \ell_0)) \ + \ \cr
 \chi_{-\frac{1}{2}-m}  (\ell_0 (1- \ell_0)) \ +
2 \psi(1+ 2m) + 2 \psi(2+ 2m) - 4 \psi(1). 
\eea
The case $m=0$ is the one considered in the first part of this section.
The series of wave functions with $(\ell_{1, m}), (\ell_{2,m}) $
allows to describe the "dressed" two-fermion eigen-states with 
anti-parallel helicity and the series of operators (\ref{operatorffm})
their interaction in the sense of the above analytic continuation
to the complex $\omega$ from integers $-2m$.
In \cite{Kirschner:1994rq} we have shown that 
the eigenvalues obtained from (\ref{eigenm}) after this
continuation describe anti-parallel helicity fermion exchange 
in $\ln s$ accuracy.

\section{Discussion}

The perturbative Regge exchanges involving not only the leading gluonic
reggeons are relevant in special semi-hard processes and contribute e.g. to the
small $x$ behaviour of flavour non-singlet, spin or chiral-odd structure
functions. The comparison of the reggeon interaction involving fermions to
the standard BFKL case  shows interesting common symmetry pattern.

Considering the generic conformal operators formulated in terms of conformal
4-point functions, determined by the weights $(\ell_1) (\ell_2)$ and a
further parameter doublet $(d)$, we have identified the particular operators
describing the one-loop perturbative QCD reggeon interaction. In all cases
the QCD reggeon interaction operator appear in the decomposition of the
generic conformal operator at the singular values of the parameter
$(d) = \pm (\ell_1- \ell_2) $. In particular the BFKL one-loop kernel is
reproduced in the dipole form \cite{Mueller:1993rr}, 
applicable to two-reggeon states
described by wave functions vanishing at coinciding points (M\"obius
representation \cite{Bartels:2004ef}). 

The shift in the conformal spin proportional to the complex angular momentum
accounting for the double-logarithmic contributions in the anti-parallel
helicity fermion exchange channel would not result in single-valued integral
kernels. We propose to describe the corresponding states and interaction 
operators by analytic continuation from a series of such objects with
half-integer spins $s_m = [\ell_m] = \pm (\frac{1}{2} +m )$.

\end{document}